\documentclass[aip,apl]{revtex4-1}

\usepackage{amsmath,amssymb,graphicx,color}

\begin{document}

\title{Tantalum nitride superconducting single-photon detectors with low cut-off energy}

\author{A. Engel}
\author{A. Aeschbacher}
\author{K. Inderbitzin}
\author{A. Schilling}
\affiliation{Physics Institute, University of Zurich, Winterthurerstr.\ 190, Zurich, Switzerland}

\author{K. Il'in}
\author{M. Hofherr}
\author{M. Siegel}
\affiliation{Institute for Micro- and Nano-Electronic Systems, Karlsruhe Institute of Technology, Hertzstr.\ 16, 76187 Karlsruhe, Germany}

\author{A. Semenov}
\author{H.-W. H\"{u}bers}
\affiliation{DLR Institute of Planetary Research, Rutherfordstr.\ 2, 12489 Berlin, Germany}

\date{\today}

\begin{abstract}
Materials with a small superconducting energy gap are expected to favor a high detection efficiency of low-energy photons in superconducting nanowire single-photon detectors. We developed a TaN detector with smaller gap and lower density of states at the Fermi energy than in comparable NbN devices, while other relevant parameters remain essentially unchanged. The observed reduction of the minimum photon energy required for direct detection is in line with model predictions of $\approx1/3$ as compared to NbN.

%This results in a reduction of the minimum photon energy required for direct detection to $\approx1/3$ as compared to NbN.
\end{abstract}


\maketitle

Superconducting nanowire single-photon detectors \cite{Semenov01,*Goltsman01} (SNSPD) are viable detectors for applications where speed is critical, both in terms of small jitter and short reset times. To date the majority of SNSPD have been made from NbN thin films due to their favorable characteristics. Superconducting NbN films with a $T_c\approx 15$~K can be made a few nanometer thin \cite{Semenov09a} and the resulting films can be structured down to strip widths of a few tens of nanometers \cite{Bartolf10, Marsili11a} without destruction of superconductivity. %These structures are very stable even in ambient conditions.
According to simple detection models, \cite{Semenov03} the threshold energy for direct detection decreases for materials with shorter thermalization time $\tau_{\text{th}}$, small electron-diffusion coefficient $D$, low density of states at the Fermi energy $N_0$ and small superconducting energy gap $\Delta$. Newer results also indicate that a large magnetic penetration depth $\lambda_L$ leading to a larger kinetic inductance $L_k\propto\lambda^2_L$ is beneficial to avoid latching into the normal conducting state.\cite{Kerman09}

All of these requirements limit the number of superconducting materials that are suitable for SNSPD, and there have been only a few publications about detectors made from alternative materials. Results on Nb-based SNSPD\cite{Engel04a,Annunziata09} highlighted the importance of $D$ and $N_0$ and confirmed the importance of $L_k$. Thin films from NbTiN offer advantages in fabrication, and the resulting SNSPD may have lower dark-count rates.\cite{Dorenbos08} Detectors made from MgB$_2$ still suffer from the low quality of ultra-thin MgB$_2$ films.\cite{Shibata10} Very recently, two reports with similar objectives have been published. In one work the use of a-W$_x$Si$_{1-x}$ with $T_c\approx3$~K resulted in increased detection efficiencies ($DE$) at lower energies.\cite{Baek11} The other study compared $DE$ of SNSPD from NbSi ($T_c\approx2$~K) with NbTiN, but although relative $DE$ increased at long wavelengths for NbSi, absolute values of $DE$ were very low. \cite{Dorenbos11}
In this letter we report results on a SNSPD made from ultra-thin TaN films. TaN is chemically and physically very similar to NbN and so are most of the relevant parameters, except for $T_c$ ($\approx6$--$10.5$~K) and the associated energy gap $\Delta$, which both are significantly smaller, and a moderate reduction of $N_0$. Any difference in detector performance can thus be linked to a change in these parameters.

Ultrathin TaN films were grown by DC reactive magnetron sputtering in an Ar/N$_2$ atmosphere on R-plane cut sapphire substrates. The sputter target was pure (99.95\%) Ta and the sapphire substrates were heated to 750$^\circ$C.
The critical temperature of the as grown films varied between $\approx10.5$~K for film thicknesses $d\geq10$~nm and $6$~K for a thickness of only $2.3$~nm. From these films SNSPD with the typical meander geometry were fabricated using electron-beam lithography and ion milling. More details about the fabrication process have been published elsewhere.\cite{Ilin11a}

The optical detector-measurements have been performed in a He-3 bath cryostat.
The temperature could be stabilized to $\approx\pm10$~mK at $5$~K and $\pm 1$~mK below $2$~K. The detector signal was transmitted to a cryogenic amplifier at the 4~K-stage and then further to a second amplifier at room temperature before fed into a $3.5$~GHz digital oscilloscope or a pulse counter. The amplifier chain had an effective bandwidth of about 40~MHz to 1.9~GHz. The bias current was applied in constant-voltage mode and passed through a series of low-pass filters. The light from a xenon discharge lamp was passed through a grating monochromator and then fed into the cryostat using a free-space setup. Although an absolute calibration of the light intensity was difficult, the lamp spectrum at the detector has been measured and the intensity was monitored during experiments to account for variations in the lamp output. The beam was slightly defocused to obtain a uniform photon-flux density over the meander area (max. $\sim 10^6$ photons $\mu$m$^{-2}$ s$^{-1}$). However, the measurements are prone to systematic errors. The given DE should therefore be taken as relative numbers. In another setup with a different TaN SNSPD an absolute DE$\approx20$\% has been determined.\cite{Ilin11a}
Complimentary resistance $R$ \emph{vs.}\ temperature $T$ measurements were performed in a \emph{Quantum Design} PPMS-9.

The detection mechanism of SNSPD relies on the conversion of the energy of the absorbed photon into elementary excitations of the superconducting film.\cite{Semenov01} Neglecting quasi-particle diffusion, one can estimate the volume of the superconducting film that switches into the normal-conducting state by equating the superconducting condensation energy of that volume to the photon energy that is converted into quasi-particle excitations, $Ad\Delta F=\zeta h\nu$, with $\Delta F$ being the free-energy density difference between the superconducting and normal states, $A$ the normal-conducting hot-spot area, $\zeta\leq1$ the conversion efficiency accounting for losses during the energy conversion process, $h$ the Planck constant and $\nu$ the photon frequency. Depending on the applied bias current $I_b$ with respect to the depairing critical current $I_c$ one can determine a minimum energy (or maximum wavelength) that can be detected.\cite{Maingault10} Taking also into account quasi-particle diffusion and the reduction of the critical-current density by excess quasi-particles one arrives at a slightly different criterion for direct detection of absorbed photons,\cite{Semenov05a}

\begin{equation}\label{Eq.Detection}
h\nu=\frac{hc}{\lambda} \geq \frac{N_0\Delta^2 wd}{\zeta}\sqrt{\pi D\tau_{th}}\left(1-\frac{I_b}{I_c}\right),
\end{equation}

with $c$ the speed of light, $\lambda$ the photon wavelength and $\tau_{th}$ the time scale of the quasi-particle multiplication process. From Eq.~\eqref{Eq.Detection} it becomes clear that the most important material parameters are $N_0$ and $\Delta$. From the analysis of $R(T)$ measurements\cite{Bartolf10} we determined all parameters relevant for the detector presented, except for the time constant $\tau_{th}\approx7$~ps, which we assumed to be similar to NbN films.\cite{Ilin98} In table \ref{Tab.Parameters} we compare our results with parameters for a reference NbN detector with almost the same cross-sectional area of the meander strip. The critical current $I_{c,\mathrm{GL}}$ given in table \ref{Tab.Parameters} is the theoretically expected depairing critical current from GL-theory at zero temperature. The experimentally achieved critical currents $I_c$ were $\approx85$\% of $I_{c,\mathrm{GL}}$.

\begin{table*}
\caption{\label{Tab.Parameters} Material and device parameters of the TaN SNSPD of this study and a reference NbN detector. $w$ is the width, $d$ the thickness, $L$ the total length of the meander and $\rho_\square$ is the square resistance of the superconducting film just above $T_c$. The superconducting energy gap has been calculated from the BCS-relation $\Delta=1.76k_BT_c$ (with the Boltzmann constant $k_B$).}
\begin{tabular}{lrrrrrrrrrr}
 & $w$ & $d$ & $L$ & $T_c$ & $N_0$ & $\Delta$ & $D$ & $I_{c,\mathrm{GL}}$ & $\xi$ & $\rho_\square$ \\
 & (nm) & (nm) & ($\mu$m) & (K) & (nm$^{-3}$eV$^{-1}$) & (meV) & (cm$^2$ s$^{-1}$) & ($\mu$A) & (nm) & ($\Omega$)\\
 \hline
 TaN & 126 & 3.9 & 71.4 & 8.16 & 44 & 1.24 & 0.6  & 22.4 & 5.5 & 590\\
 NbN & 80  & 6.0 & 120 & 13.0 & 51 & 1.98 & 0.54 & 48.2 & 4.3 & 380\\
\end{tabular}
\end{table*}

\begin{figure}
\includegraphics[width=\columnwidth,height=\textheight,keepaspectratio]{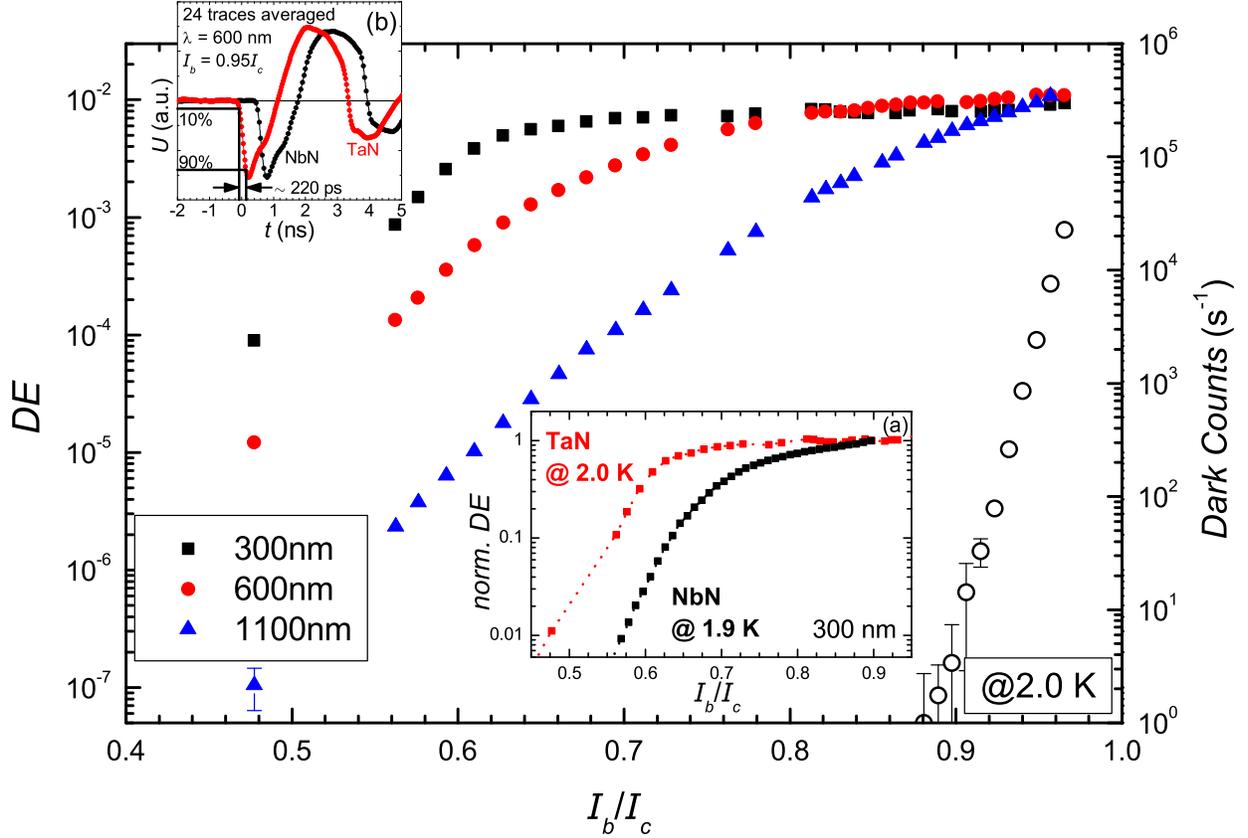}
\caption{\label{Fig.bias} Detection efficiency ($DE$) as a function of $I_b/I_c$ measured at $2.0$~K and photon wavelengths as indicated. The shift in threshold current with different photon energies is roughly linear as suggested by Eq.~\eqref{Eq.Detection}.  Open symbols show dark-count rates (right axes) as a function of bias current for $2.0$~K. Inset (a) shows a comparison between TaN and NbN detectors for $300$~nm photons and similar operating temperatures. Inset (b): Averaged and normalized traces of single-photon detection events in TaN and NbN.}
\end{figure}

Well below the critical temperature ($\lesssim0.5T_c$) and biased with a direct current of 80\% to 90\% $I_c(T)$ one can observe voltage transients that look very similar to those monitored in NbN detectors. The amplitude of the pulse varies with the applied current, the rise time of about $220$~ps (see Fig.~\ref{Fig.bias} (b)) is determined by the amplifier and oscilloscope bandwidth, the damped oscillations following the pulse are a consequence of the effective values of inductance, capacitance and resistance in the measuring circuit. In the following we assume that the same single-photon detection mechanism as for NbN SNSPD applies to TaN detectors as well.

In Fig.~\ref{Fig.bias} we present measured $DE$ as a function of the applied bias current $I_b/I_c$ for different photon wavelengths at a detector temperature of $2.0$~K. For small wavelengths, \emph{i.e.}\ high photon energies, a clearly identifiable plateau exists for bias currents above a wavelength-dependent threshold value. Below this threshold value the detection efficiency drops in an approximately exponential way with decreasing bias current. This general behavior is analogous to the typical behavior observed for NbN SNSPD. For comparison, the normalized detection efficiencies for the TaN and a reference NbN detector measured with 300 nm photons at comparable operating temperatures are plotted in the inset of Fig.~\ref{Fig.bias}. The difference in the threshold current is obvious, despite the roughly equal cross-sectional areas of the two conduction paths. The dark-count rates $R_{dc}$ are shown in the same graph as a function of bias current.

\begin{figure}
\includegraphics[width=\columnwidth,height=\textheight,keepaspectratio]{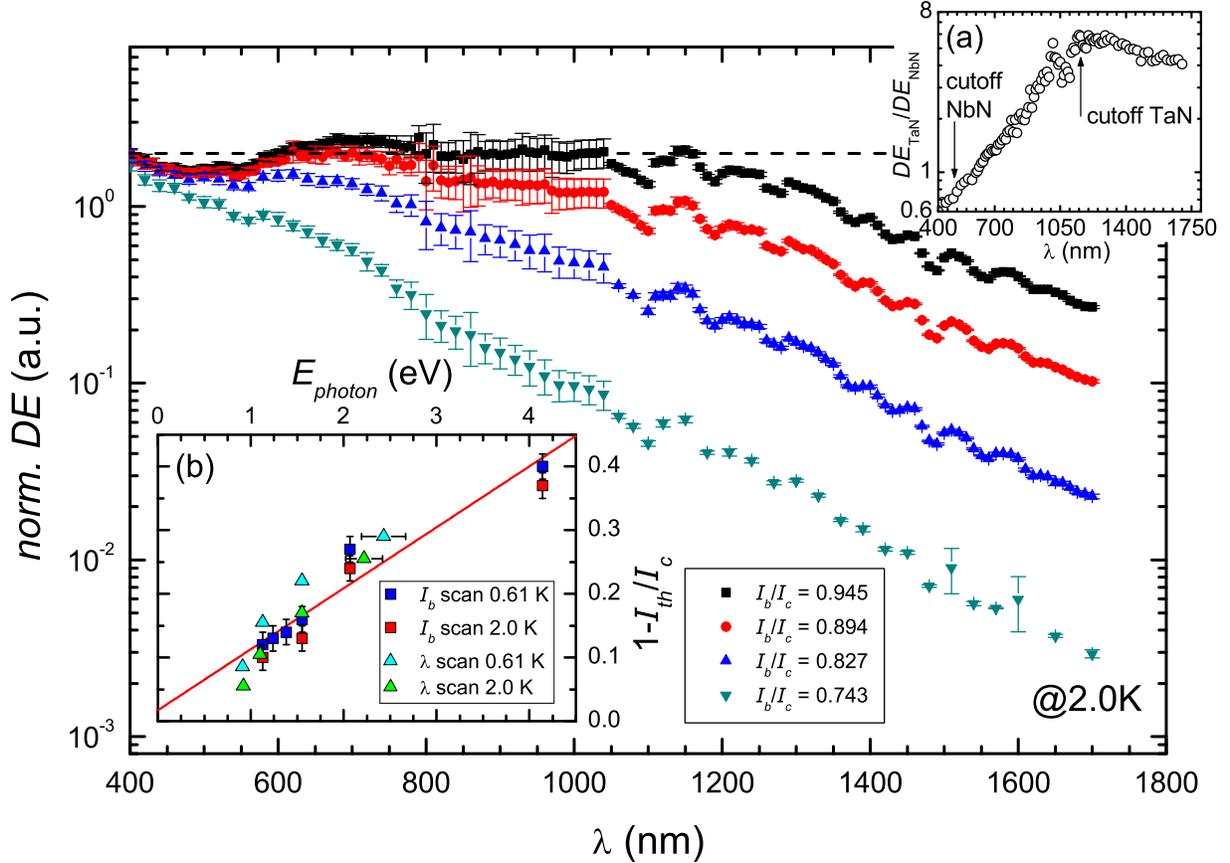}
\caption{\label{Fig.lambda} Detection efficiency ($DE$) as a function of $\lambda$ at different bias currents and $T=2.0$~K. The averaged maximum $DE$ is indicated by the horizontal dashed line. %The noise in the data is predominantly due to the light-intensity calibration.
Inset (a): Ratio of DE (DE$_\mathrm{TaN}$/DE$_\mathrm{NbN}$) \emph{vs}. $\lambda$. DE$_\mathrm{TaN}$ is significantly increased for long wavelengths and $\lambda_\mathrm{cutoff}$ is easily identified. $\lambda_\mathrm{cutoff}$ for NbN is less clear due to large systematic errors in this wavelength range. Inset (b): relation between minimum photon energy $E_{\mathrm{photon}}=h\nu$ and the threshold current $I_{th}$, error bars are estimates of the accuracy in determining the cut-off criterium. The red line is a least-square fit according to Eq.~\eqref{Eq.Detection}.}
\end{figure}

Additional measurements were also done as a function of the photon wavelength $\lambda$ for a fixed bias current. The results obtained at $T=2.0$~K are shown in Fig.~\ref{Fig.lambda} (the $DE$ show a certain repeating noise pattern that is at least partly caused by an elliptical polarization of the incident light). We normalized the $DE$ to the average $DE$ from $300$ nm to $1000$ nm obtained at the highest bias-current.

The $DE$ as a function of bias current and photon wavelength were also measured at temperatures of $0.61$~K and $4.0$~K (not shown). We have observed the same trends that were also reported for NbN\cite{Yamashita10} and Nb\cite{Annunziata09} detectors: The detector performance can be significantly improved by lowering the temperature from $4.0$ to $2.0$~K. The cut-off wavelengths are measurably longer at $2.0$~K, and $R_{dc}$ are lower by two orders of magnitude for equal $I_b/I_c$-values. For even lower temperatures ($0.61$~K), the dark-count rates are further reduced by almost two orders of magnitude, whereas we observe only a small change in the detection properties towards a lower cut-off energy.

In the inset (a) of Fig.~\ref{Fig.lambda} we plot the ratio of normalized DE for the TaN and NbN detectors \emph{vs}.\ $\lambda$ for roughly equal $I_b/I_c$. It demonstrates the significant increase in DE of the TaN SNSPD for long wavelengths and allows for an easy identification of the cutoff wavelength at around $1100$~nm.

From the data presented in Figs.~\ref{Fig.bias} and \ref{Fig.lambda} we extracted pairs of threshold bias-currents $I_{th}$ and cut-off photon wavelengths $\lambda_{max}$, which we can use to verify the detection criterion given in Eq.~\eqref{Eq.Detection}. We defined the experimental threshold for direct single-photon detection, for which the relation in Eq.~\eqref{Eq.Detection} becomes an equality, as that point, where the $DE$ reaches half the maximum $DE$. We repeated this procedure for all the $DE$-measurements as a function of $I_b$ and $\lambda$ at both $0.61$ and $2.0$~K ($4.0$~K data have been excluded, see above). %As described above, at $4.0$~K the detector characteristics are different, therefore we did not include those measurements in this analysis.
The resulting pairs of threshold currents and minimum photon energies are plotted in the inset (b) of Fig.~\ref{Fig.lambda} as $(1-I_{th}/I_c)$ \emph{vs}. $E_{min}=hc/\lambda_{max}$. According to Eq.~\eqref{Eq.Detection} these data should fall onto a single straight line through the origin, which is, within the accuracy of our data, indeed the case. Using the device parameters from Tab.~\ref{Tab.Parameters} and assuming a thermalization time $\tau_{th}=7$ ps, we can determine the conversion efficiency $\zeta\approx0.12$, which is similar to results obtained on NbN.\cite{Semenov05a}

With this value for $\zeta$ we may also calculate, using Eq.~\eqref{Eq.Detection}, the minimum photon energies required for direct detection in TaN and NbN detectors under otherwise equal operating conditions. We obtain a ratio $E_{min}(\mathrm{TaN})/E_{min}(\mathrm{NbN})\approx1/3$, which compares favorably with the observed ratio of $0.4$ to $0.5$. This good agreement also justifies our assumption of roughly equal $\tau_{th}$ in NbN and TaN.

In conclusion, we have presented results on a TaN SNSPD that showed improved detection at longer wavelengths as compared to similar sized NbN detectors. The detector performance in terms of minimum threshold-currents and cut-off wavelengths could be well described within a detection model taking into account quasi-particle multiplication and diffusion. This confirms the importance of the superconducting gap and the density of states for predicting the applicability of a certain superconducting material in SNSPD. With further improvements in TaN-film preparation and nanolithography we expect to reach $DE$ comparable to the best NbN devices without compromises in speed or jitter, but for lower photon energies. Compared to other low-gap materials recently suggested \cite{Baek11, Dorenbos11} that work best at sub-Kelvin temperatures, we identify the following advantages. Like the NbN SNSPD, the TaN based devices reach the best, nearly temperature independent performance already at about $2$~K. We also observe a significant increase in DE over NbN SNSPD in the infrared, and not only a slower decrease of DE as in NbSi compared to NbTiN. Beyond the possibility to increase the usable spectral range towards lower photon energies, TaN has also relatively short absorption lengths for X-ray photons of keV-energies.

% we achieve similar results, but at the higher temperature of $2$~K. Not only can TaN SNSPD increase the usable spectral range towards lower photon energies, TaN also has relatively short absorption lengths for X-ray photons of keV-energies. It may therefore also be a suitable material for the development of fast and sensitive X-ray detectors.

This research received support from the Swiss National Science Foundation grant No.\ 200021\_135504/1 and is supported in part by DFG Center for Functional Nanostructures under sub-project A4.3.


\begin{thebibliography}{18}%
\makeatletter
\providecommand \@ifxundefined [1]{%
 \@ifx{#1\undefined}
}%
\providecommand \@ifnum [1]{%
 \ifnum #1\expandafter \@firstoftwo
 \else \expandafter \@secondoftwo
 \fi
}%
\providecommand \@ifx [1]{%
 \ifx #1\expandafter \@firstoftwo
 \else \expandafter \@secondoftwo
 \fi
}%
\providecommand \natexlab [1]{#1}%
\providecommand \enquote  [1]{``#1''}%
\providecommand \bibnamefont  [1]{#1}%
\providecommand \bibfnamefont [1]{#1}%
\providecommand \citenamefont [1]{#1}%
\providecommand \href@noop [0]{\@secondoftwo}%
\providecommand \href [0]{\begingroup \@sanitize@url \@href}%
\providecommand \@href[1]{\@@startlink{#1}\@@href}%
\providecommand \@@href[1]{\endgroup#1\@@endlink}%
\providecommand \@sanitize@url [0]{\catcode `\\12\catcode `\$12\catcode
  `\&12\catcode `\#12\catcode `\^12\catcode `\_12\catcode `\%12\relax}%
\providecommand \@@startlink[1]{}%
\providecommand \@@endlink[0]{}%
\providecommand \url  [0]{\begingroup\@sanitize@url \@url }%
\providecommand \@url [1]{\endgroup\@href {#1}{\urlprefix }}%
\providecommand \urlprefix  [0]{URL }%
\providecommand \Eprint [0]{\href }%
\providecommand \doibase [0]{http://dx.doi.org/}%
\providecommand \selectlanguage [0]{\@gobble}%
\providecommand \bibinfo  [0]{\@secondoftwo}%
\providecommand \bibfield  [0]{\@secondoftwo}%
\providecommand \translation [1]{[#1]}%
\providecommand \BibitemOpen [0]{}%
\providecommand \bibitemStop [0]{}%
\providecommand \bibitemNoStop [0]{.\EOS\space}%
\providecommand \EOS [0]{\spacefactor3000\relax}%
\providecommand \BibitemShut  [1]{\csname bibitem#1\endcsname}%
\let\auto@bib@innerbib\@empty
%</preamble>
\bibitem [{\citenamefont {Semenov}, \citenamefont {Gol'tsman},\ and\
  \citenamefont {Korneev}(2001)}]{Semenov01}%
  \BibitemOpen
  \bibfield  {author} {\bibinfo {author} {\bibfnamefont {A.~D.}\ \bibnamefont
  {Semenov}}, \bibinfo {author} {\bibfnamefont {G.~N.}\ \bibnamefont
  {Gol'tsman}}, \ and\ \bibinfo {author} {\bibfnamefont {A.~A.}\ \bibnamefont
  {Korneev}},\ }\href@noop {} {\bibfield  {journal} {\bibinfo  {journal} {Phys.
  C}\ }\textbf {\bibinfo {volume} {351}},\ \bibinfo {pages} {349} (\bibinfo
  {year} {2001})}\BibitemShut {NoStop}%
\bibitem [{\citenamefont {Gol'tsman}\ \emph {et~al.}(2001)\citenamefont
  {Gol'tsman}, \citenamefont {Okunev}, \citenamefont {Chulkova}, \citenamefont
  {Lipatov}, \citenamefont {Semenov}, \citenamefont {Smirnov}, \citenamefont
  {Voronov}, \citenamefont {Dzardanov}, \citenamefont {Williams},\ and\
  \citenamefont {Sobolewski}}]{Goltsman01}%
  \BibitemOpen
  \bibfield  {author} {\bibinfo {author} {\bibfnamefont {G.~N.}\ \bibnamefont
  {Gol'tsman}}, \bibinfo {author} {\bibfnamefont {O.}~\bibnamefont {Okunev}},
  \bibinfo {author} {\bibfnamefont {G.}~\bibnamefont {Chulkova}}, \bibinfo
  {author} {\bibfnamefont {A.}~\bibnamefont {Lipatov}}, \bibinfo {author}
  {\bibfnamefont {A.}~\bibnamefont {Semenov}}, \bibinfo {author} {\bibfnamefont
  {K.}~\bibnamefont {Smirnov}}, \bibinfo {author} {\bibfnamefont
  {B.}~\bibnamefont {Voronov}}, \bibinfo {author} {\bibfnamefont
  {A.}~\bibnamefont {Dzardanov}}, \bibinfo {author} {\bibfnamefont
  {C.}~\bibnamefont {Williams}}, \ and\ \bibinfo {author} {\bibfnamefont
  {R.}~\bibnamefont {Sobolewski}},\ }\href@noop {} {\bibfield  {journal}
  {\bibinfo  {journal} {Appl. Phys. Lett.}\ }\textbf {\bibinfo {volume} {79}},\
  \bibinfo {pages} {705} (\bibinfo {year} {2001})}\BibitemShut {NoStop}%
\bibitem [{\citenamefont {Semenov}\ \emph {et~al.}(2009)\citenamefont
  {Semenov}, \citenamefont {G\"{u}nther}, \citenamefont {B\"{o}ttger},
  \citenamefont {H\"{u}bers}, \citenamefont {Bartolf}, \citenamefont {Engel},
  \citenamefont {Schilling}, \citenamefont {Ilin}, \citenamefont {Siegel},
  \citenamefont {Schneider}, \citenamefont {Gerthsen},\ and\ \citenamefont
  {Gippius}}]{Semenov09a}%
  \BibitemOpen
  \bibfield  {author} {\bibinfo {author} {\bibfnamefont {A.}~\bibnamefont
  {Semenov}}, \bibinfo {author} {\bibfnamefont {B.}~\bibnamefont
  {G\"{u}nther}}, \bibinfo {author} {\bibfnamefont {U.}~\bibnamefont
  {B\"{o}ttger}}, \bibinfo {author} {\bibfnamefont {H.-W.}\ \bibnamefont
  {H\"{u}bers}}, \bibinfo {author} {\bibfnamefont {H.}~\bibnamefont {Bartolf}},
  \bibinfo {author} {\bibfnamefont {A.}~\bibnamefont {Engel}}, \bibinfo
  {author} {\bibfnamefont {A.}~\bibnamefont {Schilling}}, \bibinfo {author}
  {\bibfnamefont {K.}~\bibnamefont {Ilin}}, \bibinfo {author} {\bibfnamefont
  {M.}~\bibnamefont {Siegel}}, \bibinfo {author} {\bibfnamefont
  {R.}~\bibnamefont {Schneider}}, \bibinfo {author} {\bibfnamefont
  {D.}~\bibnamefont {Gerthsen}}, \ and\ \bibinfo {author} {\bibfnamefont
  {N.~A.}\ \bibnamefont {Gippius}},\ }\href {\doibase
  10.1103/PhysRevB.80.054510} {\bibfield  {journal} {\bibinfo  {journal} {Phys.
  Rev. B}\ }\textbf {\bibinfo {volume} {80}},\ \bibinfo {pages} {054510}
  (\bibinfo {year} {2009})}\BibitemShut {NoStop}%
\bibitem [{\citenamefont {Bartolf}\ \emph {et~al.}(2010)\citenamefont
  {Bartolf}, \citenamefont {Engel}, \citenamefont {Schilling}, \citenamefont
  {Il'in}, \citenamefont {Siegel}, \citenamefont {H\"ubers},\ and\
  \citenamefont {Semenov}}]{Bartolf10}%
  \BibitemOpen
  \bibfield  {author} {\bibinfo {author} {\bibfnamefont {H.}~\bibnamefont
  {Bartolf}}, \bibinfo {author} {\bibfnamefont {A.}~\bibnamefont {Engel}},
  \bibinfo {author} {\bibfnamefont {A.}~\bibnamefont {Schilling}}, \bibinfo
  {author} {\bibfnamefont {K.}~\bibnamefont {Il'in}}, \bibinfo {author}
  {\bibfnamefont {M.}~\bibnamefont {Siegel}}, \bibinfo {author} {\bibfnamefont
  {H.-W.}\ \bibnamefont {H\"ubers}}, \ and\ \bibinfo {author} {\bibfnamefont
  {A.}~\bibnamefont {Semenov}},\ }\href {\doibase 10.1103/PhysRevB.81.024502}
  {\bibfield  {journal} {\bibinfo  {journal} {Phys. Rev. B}\ }\textbf {\bibinfo
  {volume} {81}},\ \bibinfo {pages} {024502} (\bibinfo {year}
  {2010})}\BibitemShut {NoStop}%
\bibitem [{\citenamefont {Marsili}\ \emph {et~al.}(2011)\citenamefont
  {Marsili}, \citenamefont {Najafi}, \citenamefont {Dauler}, \citenamefont
  {Bellei}, \citenamefont {Hu}, \citenamefont {Csete}, \citenamefont {Molnar},\
  and\ \citenamefont {Berggren}}]{Marsili11a}%
  \BibitemOpen
  \bibfield  {author} {\bibinfo {author} {\bibfnamefont {F.}~\bibnamefont
  {Marsili}}, \bibinfo {author} {\bibfnamefont {F.}~\bibnamefont {Najafi}},
  \bibinfo {author} {\bibfnamefont {E.}~\bibnamefont {Dauler}}, \bibinfo
  {author} {\bibfnamefont {F.}~\bibnamefont {Bellei}}, \bibinfo {author}
  {\bibfnamefont {X.}~\bibnamefont {Hu}}, \bibinfo {author} {\bibfnamefont
  {M.}~\bibnamefont {Csete}}, \bibinfo {author} {\bibfnamefont {R.~J.}\
  \bibnamefont {Molnar}}, \ and\ \bibinfo {author} {\bibfnamefont {K.~K.}\
  \bibnamefont {Berggren}},\ }\href {\doibase 10.1021/nl2005143} {\bibfield
  {journal} {\bibinfo  {journal} {Nano Letters}\ }\textbf {\bibinfo {volume}
  {11}},\ \bibinfo {pages} {2048} (\bibinfo {year} {2011})}\BibitemShut
  {NoStop}%
\bibitem [{\citenamefont {Semenov}\ \emph {et~al.}(2003)\citenamefont
  {Semenov}, \citenamefont {Engel}, \citenamefont {Il'in}, \citenamefont
  {Gol'tsman}, \citenamefont {Siegel},\ and\ \citenamefont
  {H\"{u}bers}}]{Semenov03}%
  \BibitemOpen
  \bibfield  {author} {\bibinfo {author} {\bibfnamefont {A.}~\bibnamefont
  {Semenov}}, \bibinfo {author} {\bibfnamefont {A.}~\bibnamefont {Engel}},
  \bibinfo {author} {\bibfnamefont {K.}~\bibnamefont {Il'in}}, \bibinfo
  {author} {\bibfnamefont {G.}~\bibnamefont {Gol'tsman}}, \bibinfo {author}
  {\bibfnamefont {M.}~\bibnamefont {Siegel}}, \ and\ \bibinfo {author}
  {\bibfnamefont {H.-W.}\ \bibnamefont {H\"{u}bers}},\ }\href@noop {}
  {\bibfield  {journal} {\bibinfo  {journal} {Eur. Phys. J. AP}\ }\textbf
  {\bibinfo {volume} {21}},\ \bibinfo {pages} {171} (\bibinfo {year}
  {2003})}\BibitemShut {NoStop}%
\bibitem [{\citenamefont {Kerman}\ \emph {et~al.}(2009)\citenamefont {Kerman},
  \citenamefont {Yang}, \citenamefont {Molnar}, \citenamefont {Dauler},\ and\
  \citenamefont {Berggren}}]{Kerman09}%
  \BibitemOpen
  \bibfield  {author} {\bibinfo {author} {\bibfnamefont {A.~J.}\ \bibnamefont
  {Kerman}}, \bibinfo {author} {\bibfnamefont {J.~K.~W.}\ \bibnamefont {Yang}},
  \bibinfo {author} {\bibfnamefont {R.~J.}\ \bibnamefont {Molnar}}, \bibinfo
  {author} {\bibfnamefont {E.~A.}\ \bibnamefont {Dauler}}, \ and\ \bibinfo
  {author} {\bibfnamefont {K.~K.}\ \bibnamefont {Berggren}},\ }\href {\doibase
  10.1103/PhysRevB.79.100509} {\bibfield  {journal} {\bibinfo  {journal} {Phys.
  Rev. B}\ }\textbf {\bibinfo {volume} {79}},\ \bibinfo {pages} {100509(R)}
  (\bibinfo {year} {2009})}\BibitemShut {NoStop}%
\bibitem [{\citenamefont {Engel}\ \emph {et~al.}(2004)\citenamefont {Engel},
  \citenamefont {Semenov}, \citenamefont {H\"{u}bers}, \citenamefont {Il'in},\
  and\ \citenamefont {Siegel}}]{Engel04a}%
  \BibitemOpen
  \bibfield  {author} {\bibinfo {author} {\bibfnamefont {A.}~\bibnamefont
  {Engel}}, \bibinfo {author} {\bibfnamefont {A.}~\bibnamefont {Semenov}},
  \bibinfo {author} {\bibfnamefont {H.-W.}\ \bibnamefont {H\"{u}bers}},
  \bibinfo {author} {\bibfnamefont {K.}~\bibnamefont {Il'in}}, \ and\ \bibinfo
  {author} {\bibfnamefont {M.}~\bibnamefont {Siegel}},\ }\href@noop {}
  {\bibfield  {journal} {\bibinfo  {journal} {J. Mod. Optics}\ }\textbf
  {\bibinfo {volume} {51}},\ \bibinfo {pages} {1459} (\bibinfo {year}
  {2004})}\BibitemShut {NoStop}%
\bibitem [{\citenamefont {Annunziata}\ \emph {et~al.}(2009)\citenamefont
  {Annunziata}, \citenamefont {Santavicca}, \citenamefont {Chudow},
  \citenamefont {Frunzio}, \citenamefont {Rooks}, \citenamefont {Frydman},\
  and\ \citenamefont {Prober}}]{Annunziata09}%
  \BibitemOpen
  \bibfield  {author} {\bibinfo {author} {\bibfnamefont {A.~J.}\ \bibnamefont
  {Annunziata}}, \bibinfo {author} {\bibfnamefont {D.~F.}\ \bibnamefont
  {Santavicca}}, \bibinfo {author} {\bibfnamefont {J.~D.}\ \bibnamefont
  {Chudow}}, \bibinfo {author} {\bibfnamefont {L.}~\bibnamefont {Frunzio}},
  \bibinfo {author} {\bibfnamefont {M.~J.}\ \bibnamefont {Rooks}}, \bibinfo
  {author} {\bibfnamefont {A.}~\bibnamefont {Frydman}}, \ and\ \bibinfo
  {author} {\bibfnamefont {D.~E.}\ \bibnamefont {Prober}},\ }\href {\doibase
  10.1109/TASC.2009.2018740} {\bibfield  {journal} {\bibinfo  {journal} {IEEE
  Trans. Appl. Supercon.}\ }\textbf {\bibinfo {volume} {19}},\ \bibinfo {pages}
  {327} (\bibinfo {year} {2009})}\BibitemShut {NoStop}%
\bibitem [{\citenamefont {Dorenbos}\ \emph {et~al.}(2008)\citenamefont
  {Dorenbos}, \citenamefont {Reiger}, \citenamefont {Perinetti}, \citenamefont
  {Zwiller}, \citenamefont {Zijlstra},\ and\ \citenamefont
  {Klapwijk}}]{Dorenbos08}%
  \BibitemOpen
  \bibfield  {author} {\bibinfo {author} {\bibfnamefont {S.~N.}\ \bibnamefont
  {Dorenbos}}, \bibinfo {author} {\bibfnamefont {E.~M.}\ \bibnamefont
  {Reiger}}, \bibinfo {author} {\bibfnamefont {U.}~\bibnamefont {Perinetti}},
  \bibinfo {author} {\bibfnamefont {V.}~\bibnamefont {Zwiller}}, \bibinfo
  {author} {\bibfnamefont {T.}~\bibnamefont {Zijlstra}}, \ and\ \bibinfo
  {author} {\bibfnamefont {T.~M.}\ \bibnamefont {Klapwijk}},\ }\href {\doibase
  10.1063/1.2990646} {\bibfield  {journal} {\bibinfo  {journal} {Appl. Phys.
  Lett.}\ }\textbf {\bibinfo {volume} {93}},\ \bibinfo {pages} {131101}
  (\bibinfo {year} {2008})}\BibitemShut {NoStop}%
\bibitem [{\citenamefont {Shibata}\ \emph {et~al.}(2010)\citenamefont
  {Shibata}, \citenamefont {Takesue}, \citenamefont {Honjo}, \citenamefont
  {Akazaki},\ and\ \citenamefont {Tokura}}]{Shibata10}%
  \BibitemOpen
  \bibfield  {author} {\bibinfo {author} {\bibfnamefont {H.}~\bibnamefont
  {Shibata}}, \bibinfo {author} {\bibfnamefont {H.}~\bibnamefont {Takesue}},
  \bibinfo {author} {\bibfnamefont {T.}~\bibnamefont {Honjo}}, \bibinfo
  {author} {\bibfnamefont {T.}~\bibnamefont {Akazaki}}, \ and\ \bibinfo
  {author} {\bibfnamefont {Y.}~\bibnamefont {Tokura}},\ }\href {\doibase
  10.1063/1.3518723} {\bibfield  {journal} {\bibinfo  {journal} {Appl. Phys.
  Lett.}\ }\textbf {\bibinfo {volume} {97}},\ \bibinfo {eid} {212504} (\bibinfo
  {year} {2010})}\BibitemShut {NoStop}%
\bibitem [{\citenamefont {Baek}, \citenamefont {Lita},\ and\ \citenamefont
  {Verma}(2011)}]{Baek11}%
  \BibitemOpen
  \bibfield  {author} {\bibinfo {author} {\bibfnamefont {B.}~\bibnamefont
  {Baek}}, \bibinfo {author} {\bibfnamefont {A.~E.}~\bibnamefont {Lita}}, \bibinfo {author} {\bibfnamefont{V.}~\bibnamefont {Verma}},\ and\
  \bibinfo {author} {\bibfnamefont {S.~W.}~\bibnamefont {Nam}},\ }\href {\doibase 10.1063/1.3600793} {\bibfield
  {journal} {\bibinfo  {journal} {Appl. Phys. Lett.}\ }\textbf {\bibinfo
  {volume} {98}},\ \bibinfo {pages} {251105} (\bibinfo {year}
  {2011})}\BibitemShut {NoStop}%
\bibitem [{\citenamefont {Dorenbos}\ \emph {et~al.}(2011)\citenamefont
  {Dorenbos}, \citenamefont {Forn-D\'{\i}az}, \citenamefont {Fuse},
  \citenamefont {Verbruggen}, \citenamefont {Zijlstra}, \citenamefont
  {Klapwijk},\ and\ \citenamefont {Zwiller}}]{Dorenbos11}%
  \BibitemOpen
  \bibfield  {author} {\bibinfo {author} {\bibfnamefont {S.~N.}\ \bibnamefont
  {Dorenbos}}, \bibinfo {author} {\bibfnamefont {P.}~\bibnamefont
  {Forn-D\'{\i}az}}, \bibinfo {author} {\bibfnamefont {T.}~\bibnamefont
  {Fuse}}, \bibinfo {author} {\bibfnamefont {A.~H.}\ \bibnamefont
  {Verbruggen}}, \bibinfo {author} {\bibfnamefont {T.}~\bibnamefont
  {Zijlstra}}, \bibinfo {author} {\bibfnamefont {T.~M.}\ \bibnamefont
  {Klapwijk}}, \ and\ \bibinfo {author} {\bibfnamefont {V.}~\bibnamefont
  {Zwiller}},\ }\href {\doibase 10.1063/1.3599712} {\bibfield  {journal}
  {\bibinfo  {journal} {Appl. Phys. Lett.}\ }\textbf {\bibinfo {volume}
  {98}},\ \bibinfo {eid} {251102} (\bibinfo {year} {2011})}\BibitemShut
  {NoStop}%
\bibitem [{\citenamefont {Il'in}\ \emph {et~al.}(2011)\citenamefont {Il'in},
  \citenamefont {Hofherr}, \citenamefont {Rall}, \citenamefont {Siegel},
  \citenamefont {Semenov}, \citenamefont {Engel}, \citenamefont {Inderbitzin},
  \citenamefont {Aeschbacher},\ and\ \citenamefont {Schilling}}]{Ilin11a}%
  \BibitemOpen
  \bibfield  {author} {\bibinfo {author} {\bibfnamefont {K.}~\bibnamefont
  {Il'in}}, \bibinfo {author} {\bibfnamefont {M.}~\bibnamefont {Hofherr}},
  \bibinfo {author} {\bibfnamefont {D.}~\bibnamefont {Rall}}, \bibinfo {author}
  {\bibfnamefont {M.}~\bibnamefont {Siegel}}, \bibinfo {author} {\bibfnamefont
  {A.}~\bibnamefont {Semenov}}, \bibinfo {author} {\bibfnamefont
  {A.}~\bibnamefont {Engel}}, \bibinfo {author} {\bibfnamefont
  {K.}~\bibnamefont {Inderbitzin}}, \bibinfo {author} {\bibfnamefont
  {A.}~\bibnamefont {Aeschbacher}}, \ and\ \bibinfo {author} {\bibfnamefont
  {A.}~\bibnamefont {Schilling}},\ }\href {\doibase 10.1007/s10909-011-0424-3} {\bibfield  {journal}
  {\bibinfo  {journal} {J. Low Temp. Phys. (online)}\ } (\bibinfo {year}
  {2011})}\BibitemShut {NoStop}%
\bibitem [{\citenamefont {Maingault}\ \emph {et~al.}(2010)\citenamefont
  {Maingault}, \citenamefont {Tarkhov}, \citenamefont {Florya}, \citenamefont
  {Semenov}, \citenamefont {Espiau~de Lama\"{e}stre}, \citenamefont {Cavalier},
  \citenamefont {Gol'tsman}, \citenamefont {Poizat},\ and\ \citenamefont
  {Vill\'{e}gier}}]{Maingault10}%
  \BibitemOpen
  \bibfield  {author} {\bibinfo {author} {\bibfnamefont {L.}~\bibnamefont
  {Maingault}}, \bibinfo {author} {\bibfnamefont {M.}~\bibnamefont {Tarkhov}},
  \bibinfo {author} {\bibfnamefont {I.}~\bibnamefont {Florya}}, \bibinfo
  {author} {\bibfnamefont {A.}~\bibnamefont {Semenov}}, \bibinfo {author}
  {\bibfnamefont {R.}~\bibnamefont {Espiau~de Lama\"{e}stre}}, \bibinfo
  {author} {\bibfnamefont {P.}~\bibnamefont {Cavalier}}, \bibinfo {author}
  {\bibfnamefont {G.}~\bibnamefont {Gol'tsman}}, \bibinfo {author}
  {\bibfnamefont {J.-P.}\ \bibnamefont {Poizat}}, \ and\ \bibinfo {author}
  {\bibfnamefont {J.-C.}\ \bibnamefont {Vill\'{e}gier}},\ }\href {\doibase
  10.1063/1.3374636} {\bibfield  {journal} {\bibinfo  {journal} {J. Appl.
  Phys.}\ }\textbf {\bibinfo {volume} {107}},\ \bibinfo {eid} {116103}
  (\bibinfo {year} {2010})}\BibitemShut {NoStop}%
\bibitem [{\citenamefont {Semenov}\ \emph {et~al.}(2005)\citenamefont
  {Semenov}, \citenamefont {Engel}, \citenamefont {H\"{u}bers}, \citenamefont
  {Il'in},\ and\ \citenamefont {Siegel}}]{Semenov05a}%
  \BibitemOpen
  \bibfield  {author} {\bibinfo {author} {\bibfnamefont {A.}~\bibnamefont
  {Semenov}}, \bibinfo {author} {\bibfnamefont {A.}~\bibnamefont {Engel}},
  \bibinfo {author} {\bibfnamefont {H.-W.}\ \bibnamefont {H\"{u}bers}},
  \bibinfo {author} {\bibfnamefont {K.}~\bibnamefont {Il'in}}, \ and\ \bibinfo
  {author} {\bibfnamefont {M.}~\bibnamefont {Siegel}},\ }\href {\doibase
  10.1140/epjb/e2005-00351-8} {\bibfield  {journal} {\bibinfo  {journal} {Eur.
  Phys. J. B}\ }\textbf {\bibinfo {volume} {47}},\ \bibinfo {pages} {495}
  (\bibinfo {year} {2005})}\BibitemShut {NoStop}%
\bibitem [{\citenamefont {Il'in}\ \emph {et~al.}(1998)\citenamefont {Il'in},
  \citenamefont {Milostnaya}, \citenamefont {Verevkin}, \citenamefont
  {Gol'tsman}, \citenamefont {Gershenzon},\ and\ \citenamefont
  {Sobolewski}}]{Ilin98}%
  \BibitemOpen
  \bibfield  {author} {\bibinfo {author} {\bibfnamefont {K.~S.}\ \bibnamefont
  {Il'in}}, \bibinfo {author} {\bibfnamefont {I.~I.}\ \bibnamefont
  {Milostnaya}}, \bibinfo {author} {\bibfnamefont {A.~A.}\ \bibnamefont
  {Verevkin}}, \bibinfo {author} {\bibfnamefont {G.~N.}\ \bibnamefont
  {Gol'tsman}}, \bibinfo {author} {\bibfnamefont {E.~M.}\ \bibnamefont
  {Gershenzon}}, \ and\ \bibinfo {author} {\bibfnamefont {R.}~\bibnamefont
  {Sobolewski}},\ }\href@noop {} {\bibfield  {journal} {\bibinfo  {journal}
  {Appl. Phys. Lett.}\ }\textbf {\bibinfo {volume} {73}},\ \bibinfo {pages}
  {3938} (\bibinfo {year} {1998})}\BibitemShut {NoStop}%
\bibitem [{\citenamefont {Yamashita}\ \emph {et~al.}(2010)\citenamefont
  {Yamashita}, \citenamefont {Miki}, \citenamefont {Qiu}, \citenamefont
  {Fujiwara}, \citenamefont {Sasaki},\ and\ \citenamefont
  {Wang}}]{Yamashita10}%
  \BibitemOpen
  \bibfield  {author} {\bibinfo {author} {\bibfnamefont {T.}~\bibnamefont
  {Yamashita}}, \bibinfo {author} {\bibfnamefont {S.}~\bibnamefont {Miki}},
  \bibinfo {author} {\bibfnamefont {W.}~\bibnamefont {Qiu}}, \bibinfo {author}
  {\bibfnamefont {M.}~\bibnamefont {Fujiwara}}, \bibinfo {author}
  {\bibfnamefont {M.}~\bibnamefont {Sasaki}}, \ and\ \bibinfo {author}
  {\bibfnamefont {Z.}~\bibnamefont {Wang}},\ }\href {\doibase
  10.1143/APEX.3.102502} {\bibfield  {journal} {\bibinfo  {journal} {Appl.
  Phys. Express}\ }\textbf {\bibinfo {volume} {3}},\ \bibinfo {pages}
  {102502} (\bibinfo {year} {2010})}\BibitemShut {NoStop}%
\end{thebibliography}
\end{document}